\documentclass[aps,prd,reprint,floatfix,nofootinbib,superscriptaddress]{revtex4-1}

\usepackage[T1]{fontenc}
\usepackage[utf8]{inputenc}
\usepackage{graphicx}
\usepackage[usenames,dvipsnames]{xcolor}
\usepackage{booktabs}
\usepackage{mathtools}
\usepackage{amssymb}
\usepackage{lmodern}
\usepackage{siunitx}
\usepackage{slashed}
\usepackage[final]{microtype}
\usepackage{hyperref}

\hypersetup{%
	pdftitle = {Locating the critical endpoint of QCD: mesonic backcoupling effects},
	pdfauthor = {Pascal J. Gunkel, Christian S. Fischer},
	bookmarksopen = true,
	colorlinks = true,
	linkcolor = blue!50!black,
	citecolor = blue!50!black,
	urlcolor = blue!50!black
}

\newcolumntype{d}[1]{D{.}{.}{#1}}

\newcommand*{\+}{\hspace*{.08335em}}

\newcommand*{\beq}{\begin{equation}}
\newcommand*{\eeq}{\end{equation}}
\newcommand*{\Nf}{N_{\textup{f}}}
\newcommand*{\Nc}{N_{\textup{c}}}
\newcommand*{\OO}{\operatorname{O}}
\newcommand*{\ZZ}{\operatorname{Z}}
\newcommand*{\dd}{\mathrm{d}}
\newcommand*{\ii}{\mathrm{i}}
\newcommand*{\upB}{\textup{B}}

\newcommand*{\ups}{\textup{s}}
\newcommand*{\upq}{\textup{q}}

\newcommand*{\muq}{\mu_{\upq}}
\newcommand*{\muB}{\mu_{\upB}}
\newcommand*{\Tc}{T_{\textup{c}}}

\DeclareMathOperator*{\SumInt}{%
	\mathchoice%
	{\ooalign{$\displaystyle\sum$\cr\hidewidth$\displaystyle\int$\hidewidth\cr}}
	{\ooalign{\raisebox{.14\height}{\scalebox{.7}{$\textstyle\sum$}}\cr\hidewidth$\textstyle\int$\hidewidth\cr}}
	{\ooalign{\raisebox{.2\height}{\scalebox{.6}{$\scriptstyle\sum$}}\cr$\scriptstyle\int$\cr}}
	{\ooalign{\raisebox{.2\height}{\scalebox{.6}{$\scriptstyle\sum$}}\cr$\scriptstyle\int$\cr}}
}

\DeclareMathOperator*{\argmax}{arg\,max}
\DeclareMathOperator{\Tr}{\textup{Tr}}
\DeclarePairedDelimiterX{\expval}[1]{\langle}{\rangle}{#1}

\DeclareSIUnit{\eV}{\electronvolt}
\DeclareSIUnit{\MeV}{\mega\eV}
\DeclareSIUnit{\GeV}{\giga\eV}

\begin{document}

\title{Locating the critical endpoint of QCD: mesonic backcoupling effects}

\author{Pascal J.~Gunkel}
\email{pascal.gunkel@physik.uni-giessen.de}
\affiliation{%
	Institut f\"ur Theoretische Physik, %
	Justus-Liebig-Universit\"at Gie\ss{}en, %
	35392 Gie\ss{}en, %
	Germany%
}
\affiliation{%
	Helmholtz Forschungsakademie Hessen f\"ur FAIR (HFHF), %
	GSI Helmholtzzentrum\\
	f\"ur Schwerionenforschung, Campus Gie{\ss}en, %
	35392 Gie{\ss}en, Germany%
}

\author{Christian S.~Fischer}
\email{christian.fischer@theo.physik.uni-giessen.de}
\affiliation{%
	Institut f\"ur Theoretische Physik, %
	Justus-Liebig-Universit\"at Gie\ss{}en, %
	35392 Gie\ss{}en, %
	Germany%
}
\affiliation{%
	Helmholtz Forschungsakademie Hessen f\"ur FAIR (HFHF), %
	GSI Helmholtzzentrum\\
	f\"ur Schwerionenforschung, Campus Gie{\ss}en, %
	35392 Gie{\ss}en, Germany%
}

\begin{abstract}
We study the effects of pion and sigma meson backcoupling on the chiral order parameters and the QCD phase diagram
and determine their effect on the location of the chiral critical endpoint. To this end, we solve a coupled 
set of truncated Dyson--Schwinger equations for Landau gauge quark and gluon propagators with $\Nf=2+1$ 
dynamical quark flavors and explicitly backcoupled mesons. The corresponding meson bound-state properties and the 
quark-meson Bethe--Salpeter vertices are obtained from their homogeneous Bethe--Salpeter equation. We find chiral-restoration effects of the pion and/or sigma meson backcoupling and observe a (small) shift of the critical endpoint towards smaller chemical potentials. The curvature of the chiral crossover line decreases. Our results indicate
that the location of the critical endpoint in the phase diagram is mainly determined by the microscopic degrees of 
freedom of QCD (in contrast to its critical properties).
\end{abstract}

\maketitle

% ==============================================================================
\section{\label{sec:intro}Introduction}
% ==============================================================================

The phase structure of QCD at finite chemical potential is probed in heavy-ion-collision 
experiments at RHIC/BNL \cite{Bzdak:2019pkr} and HADES (FAIR Phase-0) \cite{Salabura:2020tou}, 
as well as the future CBM/FAIR experiment \cite{Friman:2011zz}. An important goal of these 
experiments is to provide answers to the quest of the existence, the location, and the 
properties of a chiral critical endpoint (CEP).

Theoretical approaches to QCD agree with each other that no such CEP may be 
found in the region of the temperature--baryon-chemical-potential plane $(T, \muB)$ with $\muB\+/\+T < 2.5$.
This region is excluded by recent studies on the lattice, see, e.g., Refs.~\cite{Bazavov:2018mes,Borsanyi:2020fev} 
and references therein, as well as studies using functional methods \cite{Isserstedt:2019pgx,Fu:2019hdw,Gao:2020fbl}.
Beyond this region, errors in lattice extrapolations accumulate rapidly and no definite 
statements can be made. On the other hand, functional approaches, i.e., approaches via Dyson--Schwinger equations (DSE)
and/or the functional renormalisation group (FRG), do in principle allow for a mapping of the whole 
QCD phase diagram but inherently depend on approximations and truncations necessary to make 
the equations tractable. 

These truncations are necessary due to the infinite hierarchy inherent in the functional approach.
Equations governing the behavior of $n$-point functions do depend on ($n+1$)-point functions and,
in some cases, even ($n+2$)-point functions. A systematic way to address the quality of truncations 
is to work order by order in a field expansion. One starts by solving the equations for the
two-point functions (i.e., propagators) of the theory assuming ans\"atze for the higher $n$-point 
functions using guiding principles such as perturbation theory (at large momenta) and Slavnov--Taylor 
identities (at small momenta). In a next step, one also solves for the equations of the three-point
functions and so on. In vacuum QCD, this program has progressed to include all primitively divergent
$n$-point functions, i.e., all QCD propagators and vertices that appear in the QCD Lagrangian, see
Refs.~\cite{Braun:2014ata,Williams:2015cvx,Cyrol:2017ewj,Aguilar:2019uob,Huber:2020keu,Gao:2021wun}.
Direct and systematic comparison with corresponding lattice calculations of these Green's functions
suggest that truncations on this level deliver quantitatively accurate results. Consequently,
spectra of mesons and glueballs calculated from such truncations are correct on a quantitative 
level \cite{Williams:2015cvx,Huber:2020ngt}. A corresponding calculation of the spectrum of baryons
that is based on insights gained from such truncations is also in agreement with experiment \cite{Eichmann:2016hgl}.

At finite temperature, truncations applied so far have not yet reached this stage of sophistication, 
see, e.g., Ref.~\cite{Fischer:2018sdj} for a recent review. While propagators have been determined from 
their Dyson--Schwinger and FRG equations \cite{Fischer:2013eca,Fischer:2014ata,Eichmann:2015kfa,Fu:2016tey,Isserstedt:2019pgx,Fu:2019hdw,Braun:2020ada,Gao:2020qsj,Gao:2020fbl}, 
the corresponding vertices have not yet been determined with comparable precision as in the vacuum. 
This is true in particular for the quark-gluon vertex, which is the crucial element that couples the
Yang--Mills sector of QCD with its quark sector. Consequently, recent attention has focused on the
details of the medium fluctuations of this vertex and their effect on the location of the CEP. 
In the DSE framework, Ref.~\cite{Gao:2020fbl} explored effects due to nonprimitively-divergent vertex 
structures, while in Ref.~\cite{Eichmann:2015kfa} effects due to virtual loops 
containing off-shell baryons have been discussed. Furthermore, in the FRG-QCD framework, mesonic medium 
effects have been taken into account in Refs.~\cite{Fu:2016tey,Fu:2019hdw,Braun:2020ada} and are
naturally present in quark-meson type models, see e.g. \cite{Schaefer:2006ds,Schaefer:2007pw,Skokov:2010wb,Rennecke:2016tkm}. 
Due to the inherent complementarity of truncations in the DSE and FRG frameworks, it is highly desirable to 
complement these studies by a corresponding one in the DSE approach. This is the purpose of the present 
work. 

The paper is organized as follows. In Sec.~\ref{sec:truncation}, we discuss the details of our truncation
scheme and specify how we deal with the meson fluctuations. In Sec.~\ref{sec:effect_chiral_order_parameter}, we
then study the influence of these fluctuations on the chiral order parameters at zero and finite temperature.
In Sec.~\ref{sec:effect_QCD_chiral_phase_diagram}, we discuss the resulting phase diagram of QCD before
we conclude in Sec.~\ref{sec:summary}.

\begin{figure*}[t!]
	\centering
	\includegraphics[width=0.65\textwidth]{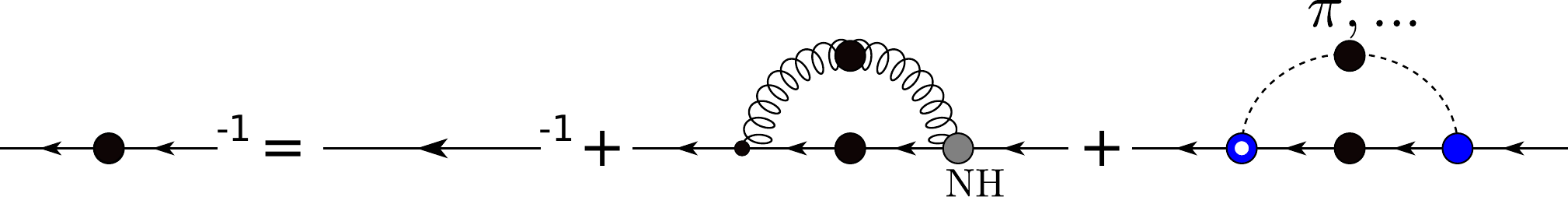} \\[1em]
	\includegraphics[width=0.45\textwidth]{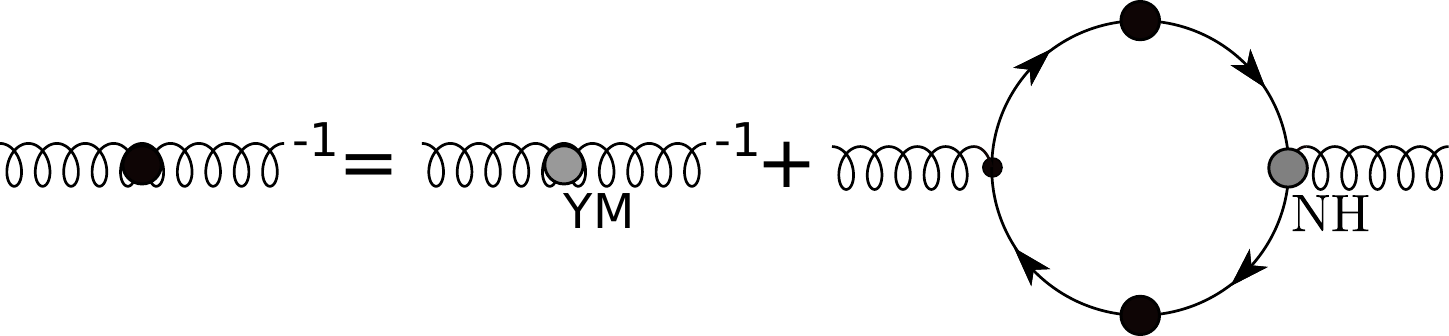}
	\hfill
	\includegraphics[width=0.45\textwidth]{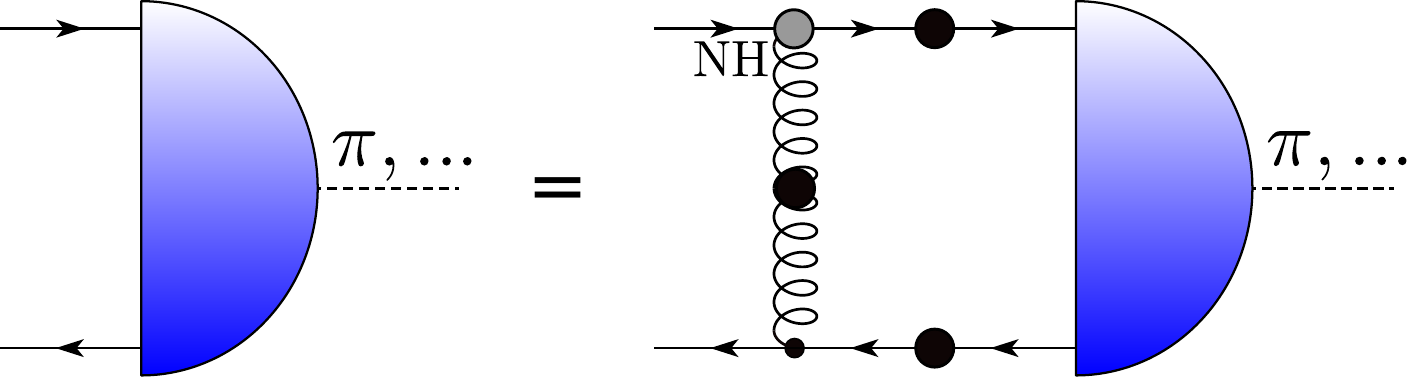}
	\caption{\label{fig:truncated_coupled_DSE_set}Truncated DSEs for the quark propagator (top) and gluon propagator (bottom left),
	and truncated Bethe--Salpeter equation (BSE) (bottom right). Quark, gluon, and meson propagators are denoted as solid, curly, and dashed lines, respectively.
	The intersection of two quarks and a gluon or a meson represent a quark-gluon or a Bethe--Salpeter vertex, respectively. Dressed quantities are indicated 
	by big full dots; the remaining ones are bare. The signs and prefactors are absorbed into the diagrams.}	
	\centering\vspace*{7mm}
	\hspace*{8mm}\includegraphics[width=0.35\textwidth]{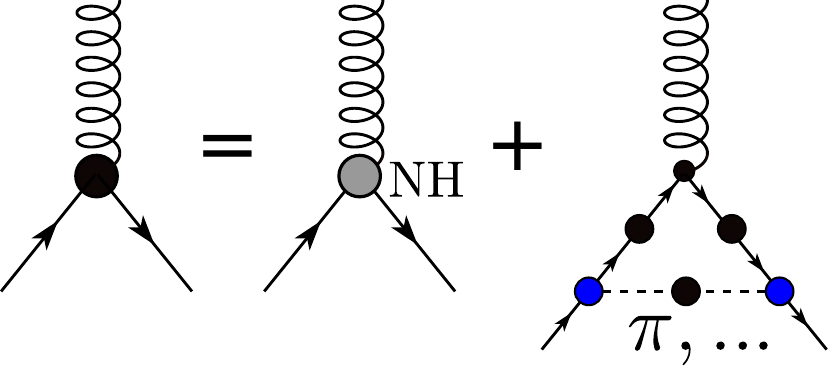}
	\hfill
	\includegraphics[width=0.45\textwidth]{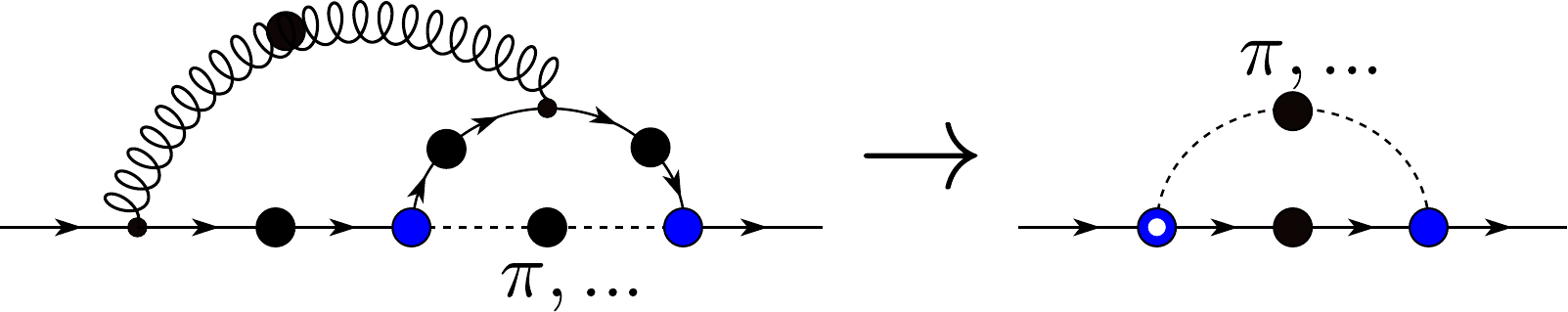}\hspace*{8mm}
	\caption{\label{fig:vertex_approximations} Left: Separation of the full quark-gluon vertex into nonhadronic (NH) contributions (first term) and lowest-order mesonic contributions resulting from a resonance expansion of the quark-antiquark scattering kernel. For the mesonic contribution, an one-meson exchange (second term) is considered. Right: Approximation of the meson-backcoupling quark self-energy resulting from the insertion of the separation of the full quark-gluon vertex into the quark DSE. The blue dot with a white center represents an effective Bethe--Salpeter vertex, which will be discussed in the text. The remaining components are defined in the same way as in Fig.~\ref{fig:truncated_coupled_DSE_set}.}	
\end{figure*}

% ==============================================================================
\section{\label{sec:truncation}Truncation}
% ==============================================================================
%
The dressed quark propagator at finite quark chemical potential $\mu_\textup{q}^f$ and temperature $T$ for the quark flavor $f$ can be represented by
\beq
\label{eq:quark_propagator}
S^{-1}_f(p)=\ii\+\vec{\slashed{p}}\+A_f(p)+\ii\+\tilde{\omega}_p^f\gamma_4\+C_f(p)+B_f(p)
\eeq
with the four-momentum $p=(\vec{p},\tilde{\omega}_p^f)$, the Matsubara frequency $\tilde{\omega}_p^f=\omega_p+\ii\+\muq^f$ 
including the quark chemical potential\footnote{We set the isospin $\mu_\textup{I}$ and strange quark $\muq^\ups$ 
chemical potential to zero, implying the relation $3\+\muq^\ell=\muB$ between quark and baryon chemical 
potential.}, and the quark dressing functions $A_f$, $B_f$, and $C_f$ that encode the nontrivial momentum dependence of the propagator. 
Together with the gluon propagator, we obtain the quark propagator from a coupled set of truncated DSEs shown 
in Fig.~\ref{fig:truncated_coupled_DSE_set}. 

The new element that is different from previous finite-temperature studies within the DSE framework is the quark-meson loop appearing in the
quark DSE. It arises from a specific diagram in the DSE for the quark-gluon vertex that involves a four-quark kernel in pole approximation, 
shown in the left diagram of Fig.~\ref{fig:vertex_approximations}. This diagram provides contributions to all tensor components of the 
quark-gluon vertex \cite{Fischer:2007ze}. In the quark DSE, the resulting two-loop diagram can be simplified to a 
one-loop diagram using the homogenuous BSE as shown in the left diagram of Fig.~\ref{fig:vertex_approximations}, see Ref.~\cite{Fischer:2007ze} 
for details. The effect of this specific contribution to the quark-gluon interaction
has been studied in a number of works at zero temperature/chemical potential including a discussion of the analytic structure
of the quark propagator \cite{Fischer:2008sp}, a discussion of its effect onto the meson spectrum \cite{Fischer:2008wy}, and
an exploratory study of meson-cloud effects in baryons \cite{Sanchis-Alepuz:2014wea}. In all these studies, it has been noted that
meson-backcoupling effects typically provide contributions of the order of 10--20\,\% as compared with other components of the
quark-gluon interaction. 

At finite temperature, however, these contributions may become dominant due to universality. This happens in the vicinity of 
the critical temperature of the second-order phase transition in the chiral limit of vanishing quark masses\footnote{An explicit 
study of this limit within the DSE framework can be found in Ref.~\cite{Fischer:2011pk}.} and also at finite quark masses close 
to the CEP. It is, however, clear that the critical region around the CEP where these fluctuations are large is actually quite 
small \cite{Schaefer:2006ds}, and therefore it is not clear to what extent the meson fluctuations are able to influence the
location of the CEP. A quantitative study of this effect is the purpose of this work. Preliminary work in this direction has been
discussed in Ref.~\cite{Lucker:2013uya}. Here, we improve upon this study by taking explicit information on the 
Bethe--Salpeter wave functions of the mesons from their BSEs into account. 

Before we specify the details of the mesonic part of the quark DSE, let us briefly summarize our treatment of the other diagram including the
gluon. All technical details have been published elsewhere \cite{Fischer:2012vc,Eichmann:2015kfa} and shall not be repeated here in order to
keep the paper concise and to the point. Let us start with the gluon. In the DSE for the gluon, all diagrams involving only Yang--Mills propagators
and vertices have been replaced by an inverse propagator that is taken from quenched lattice QCD \cite{Fischer:2010fx,Maas:2011ez}. This procedure 
ensures that all temperature fluctuations of the Yang--Mills diagrams are taken into account. Quark-loop effects in these Yang--Mills diagrams, 
however, are neglected. However, we take into account the explicit quark-loop in the gluon DSE, which contains $\Nf=2+1$ quark flavors. 
The backcoupling of the quarks to the gluon is performed using an ansatz for the quark-gluon vertex that is given by 
\begin{widetext}
\begin{align}
	\Gamma_\mu^f(p,q;k)_\textup{NH} &= \gamma_\mu \+ \Gamma(p^2,q^2,k^2)_\textup{NH} 
	\left(\delta_{\mu 4}\frac{C_f(q)+C_f(p)}{2} + (1 - \delta_{\mu 4}) \frac{A_f(q)+A_f(p)}{2} \right) ,
	\label{vertex1}
	\\[0.5em]
	\Gamma(p^2,q^2,k^2)_\textup{NH} &= \frac{d_1}{d_2+y}
	+ \frac{y}{\Lambda^2 + y}
	\left(\frac{\alpha(\nu) \+ \beta_0}{4\pi} \ln(y / \Lambda^2 + 1) \right)^{2\delta} \label{vertex2}
\end{align}
\end{widetext}
with quark momenta $p$ and $q$ and gluon momentum $k$. The squared-momentum variable $y$ is identified with the gluon momentum $k^2$ 
in the quark DSE and with the sum of the two squared quark momenta $q^2+p^2$ in the quark loops of the gluon DSE to ensure
multiplicative renormalizability. Medium effects in the leading $\gamma_\mu$ part of the vertex are taken into account by splitting into longitudinal 
and transverse parts with respect to the heat-bath vector $u=(0,0,0,1)$. The corresponding dressing functions $A_f$ and $C_f$
depend explicitly on temperature and chemical potential and stem from the quark propagator; cf.~Eq.~\eqref{eq:quark_propagator}. Their appearance is dictated 
by the Abelian part of the Slavnov--Taylor identity of the vertex. Its non-Abelian part is taken into account by an infrared-enhanced function $\Gamma(p^2,q^2,k^2)_\textup{NH}$ that also accounts for the correct ultraviolet running of the vertex. Both scales 
$\Lambda= \SI{1.4}{\GeV}$ and $d_2 = \SI{0.5}{\GeV\squared}$ are fixed such that they match the corresponding scales in the gluon lattice data. $\alpha(\nu)=0.3$ is the running coupling at
a scale fixed by the quenched gluon from the lattice.
The anomalous dimension is $\delta=-9 \Nc/(44 \Nc - 8 \Nf)$ and $\beta_0=(11\Nc-2\Nf)/3$. The only free parameter 
of the interaction is the vertex strength $d_1$, which has been adapted to pseudocritical chiral transition temperature (at $\muB=0$) determined on the 
lattice. This results in $d_1 = \SI{4.6}{\GeV\squared}$ for the quenched theory \cite{Fischer:2010fx} and $d_1 = \SI{8.49}{\GeV\squared}$ 
for the theory with $\Nf=2+1$ quark flavors \cite{Isserstedt:2019pgx}. 
 
The same quark-gluon vertex appears in the gluonic part of the quark DSE and, because of the axial Ward--Takahashi identity, also 
in the quark-antiquark interaction kernel of the meson BSE. This has the potential to complicate matters considerably, since the presence 
of the quark dressing functions $A_f$ and $C_f$ in the vertex needs to be taken into account carefully in the construction of the
kernel, see, e.g., \cite{Heupel:2014ina}. In order to simplify matters, we will resort to a truncation that has been explored already in Ref.~\cite{Gunkel:2019xnh,Gunkel:2020wcl} and use the $\OO(4)$-symmetric vertex 
\beq\label{vertexa}
	\Gamma_\mu^f(p, q; k)_\textup{NH} = Z_2^f \gamma_\mu \+ \Gamma(p^2,q^2,k^2)_\textup{NH}
\eeq
with $Z_2^f$ being the quark wave function renormalization constant and with a different parameter $d_1^{\+\upq}$ in $\Gamma(p^2,q^2,k^2)$ in the quark DSE and the meson BSE ($d_2$ and $\Lambda$ remain unchanged).
The axial Ward--Takahashi identity is then satisfied trivially. In order to account for the missing 
interaction strength due to the omission of the quark dressing functions, the infrared-strength parameter within the expression (\ref{vertexa})
needs to be adapted. The corresponding values $d_1^{\+\upq}$ for different setups are discussed below in Sec.~\ref{sec:effect_chiral_order_parameter}.
\begin{figure*}[t!]
	\centering
	\includegraphics[width=0.48\textwidth]{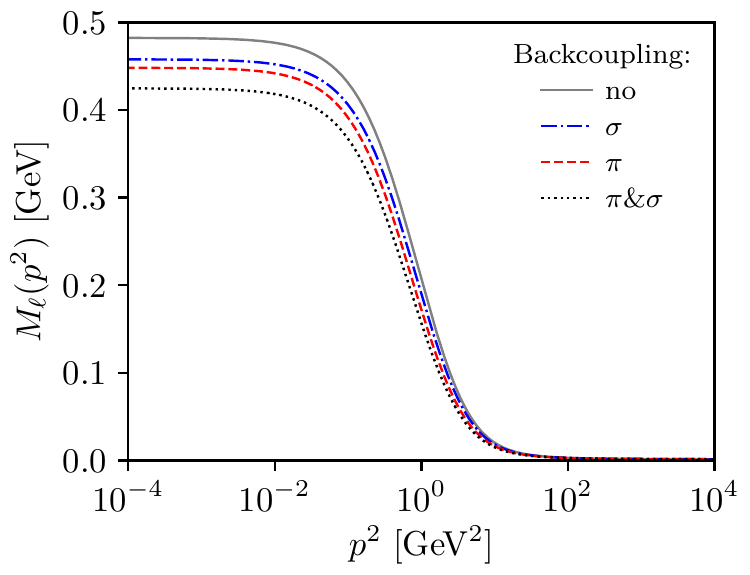}
	\hfill
	\includegraphics[width=0.48\textwidth]{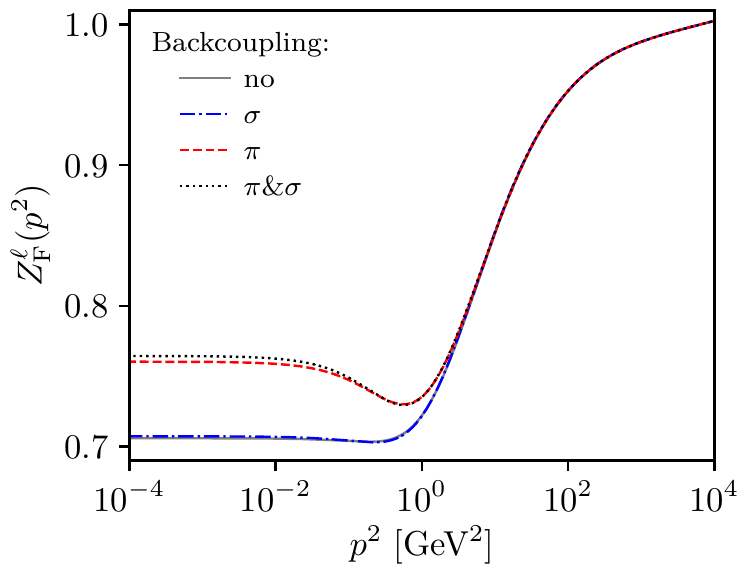}
	\caption{\label{fig:bc_dependence_quark_df}Dynamical quark mass (left) and corresponding quark wave function renormalization function $Z_\textup{F}^f(p)=1/A_f(p)$ (right) for light quarks and different numbers of backcoupled mesons in vacuum. The results are obtained 
		with a fixed strength parameter $d_1^{\+\textup{q}}=\SI{12.85}{\GeV\squared}$ \cite{Gunkel:2019xnh} in order to make the effects of the additional diagrams visible.}
\end{figure*}

We now come back to the meson diagram in the quark DSE. Following Ref.~\cite{Fischer:2008wy}, we calculate the meson-backcoupling part 
of the quark self-energy $\Sigma_f^\textup{M}$ via
\begin{align}
\label{eq:meson_bc_quark_self_energy}
\Sigma_f^\textup{M}(p)
&=
-\delta_{f\ell} \sum_{X} \SumInt_q \frac{D_X(P)}{2} \,
\Bigl[ \bar{\Gamma}_X^0 \+ S_f(q) \+ \hat{\Gamma}_X(P,l)
\nonumber
\\
&\phantom{=\;} +
\bar{\Gamma}_X^0 \+ S_f(q) \+ \hat{\Gamma}_X(-P,l) \Bigr]
\end{align}
with the shorthand notation $\SumInt_q=T\sum_{\omega_q}\int \dd^3q/(2\pi)^3$ and the on-shell total $P^2=-m_X^2$ and relative $l=\frac{p+q}{2}$ momentum of the meson. This part of the self-energy contains a sum over all 
considered mesons. We restrict ourself to the isovector pions and the isoscalar sigma meson, i.e., $X\subset\{\pi_\pm,\pi_0,\sigma\}$,
because we expect these to have the biggest influence on the QCD phase diagram. Since we work in the isospin-symmetric limit, the 
three different pions are mass degenerate and sum up to a overall flavor prefactor of $3/2$. The sigma meson on the other hand comes with a overall flavor prefactor of $1/2$.
The delta function $\delta_{f\ell}$ indicates that the pion and the sigma do not couple to the strange quark and are therefore absent 
in the strange-quark DSE. A generalization of the framework to include mesons with open and hidden strangeness, too, is straightforward but tedious and postponed to a future work. The arithmetic mean of the two terms in the square bracket is necessary to satisfy the
axial Ward--Takahashi identity and therefore preserves the Goldstone nature of the pion. 

Whereas one quark-meson vertex is given by the Bethe--Salpeter amplitude, the effective other one is taken bare \cite{Fischer:2008sp}. The corresponding 
charge-conjugated bare Bethe--Salpeter vertex is given by
\beq
\label{eq:bare_BS_vertex}
\bar{\Gamma}_X^0 =
\begin{cases}
Z_2 \+ \gamma_5 & \textup{for~} X=\pi \, , \\[0.5em]
Z_2 \+ \openone & \textup{for~} X=\sigma \, .
\end{cases}
\eeq
The back-coupling term further depends on the meson propagator 
$D_X(P)$, which is discussed in Refs.~\cite{Son:2001ff,Son:2002ci,Fischer:2011pk} and given by
\beq
\label{eq:meson_propagator}
D_X(P)=\frac{1}{P_4^2+u_X^2(\vec{P}^2 + m_X^2)}\,.
\eeq
The meson velocity $u_X=f_X^{\+\textup{s}} / f_X^{\+\textup{t}}$ is given by the ratio of the 
spatial and temporal meson decay constants $f_X^{\+\textup{s}}$ and $f_X^{\+\textup{t}}$, respectively, and is
approximated by $u_X=1$ in this work.

The last quantity to define is the normalized Bethe--Salpeter amplitude (BSA) $\hat{\Gamma}_X(P,l)$. In medium,
we use  the same tensor decomposition for the BSA of pseudo-scalar ($X=\textup{P}$) and scalar ($X=\textup{S}$) mesons as in Ref.~\cite{Gunkel:2020wcl}:
\begin{align}
\label{eq:BSA_decomposition}
\Gamma_\textup{P}(P,l)&=\gamma_5 \+ \bigl[E_\textup{P}(P,l)-\ii\+\vec{\slashed{l}}P\cdot l \, G_\textup{P}^{\+\textup{s}}(P,l)-\ii\+\gamma_4I_\textup{P}(P,l)\bigr]\,, \\
\Gamma_\textup{S}(P,l)&=E_\textup{S}(P,l)-\ii\+\vec{\slashed{l}}G_\textup{S}^{\+\textup{s}}(P,l)-\ii\+\gamma_4I_\textup{S}(P,l)\,.
\end{align}
The dressed BSAs depend on the relative momentum $l=(p+q)/2$ and the off-shell total momentum $P=q-p$, implying a symmetric momentum partitioning.
Note that strictly speaking, the amplitudes obtained from the homogenous BSE are well-defined only for on-shell mesons, whereas in the
quark DSE we need the corresponding off-shell quantities. These can be extracted from an inhomogeneous BSE and are almost identical
to the on-shell amplitudes for the momenta relevant in the quark DSE. 

In the preliminary study discussed in Ref.~\cite{Lucker:2013uya}, the meson backcoupling in medium was calculated with mesons approximated 
by generalized Goldberger--Treiman-like relations. In this work, we will resolve the meson-backcoupling effects with BSAs explicitly calculated in Ref.~\cite{Gunkel:2020wcl} 
from the homogeneous BSE. These solutions incorporate important chemical-potential effects in the BSA 
that are mandatory to preserve the Silver-Blaze property of QCD. They do not, however, include effects due to temperature fluctuations in
the meson BSE. Again, these need to be included in future work. Having outlined the formalism, we proceed with discussing our results
in the next two sections.

% ==============================================================================
\section{\label{sec:effect_chiral_order_parameter}Effect on the chiral order parameters}
% ==============================================================================

In this section, we study the effect of the mesonic backcoupling on the chiral order parameters. As chiral order parameters we consider the dynamical quark mass $M_f(p)=B_f(p)/A_f(p)$ and the regularized quark condensate
\begin{align}
\label{eq:regularized_quark_condensate}
\Delta_{\ell\ups}(T,\muB)&=\expval{\bar{\Psi}\Psi}_\ell(T,\muB)-\frac{Z_m^\ell m_\upq^\ell}{Z_m^\ups m_\upq^\ups}\expval{\bar{\Psi}\Psi}_\ups(T,\muB)\, ,\\
\expval{\bar{\Psi}\Psi}_f(T,\muB)&=-Z_2^fZ_m^f \+ \Nc \SumInt_q \Tr\bigr[S_f(q)\bigl]
\end{align}
with $Z_m^f$ and $m_\upq^f$ denoting the quark mass renormalization constant and the renormalized quark mass of the quark flavor $f$, respectively. 
The regularization prescription ensures that the quadratic divergence proportional to the quark mass drops out. 
\begin{figure}[t!]
\centering
\includegraphics[width=0.48\textwidth]{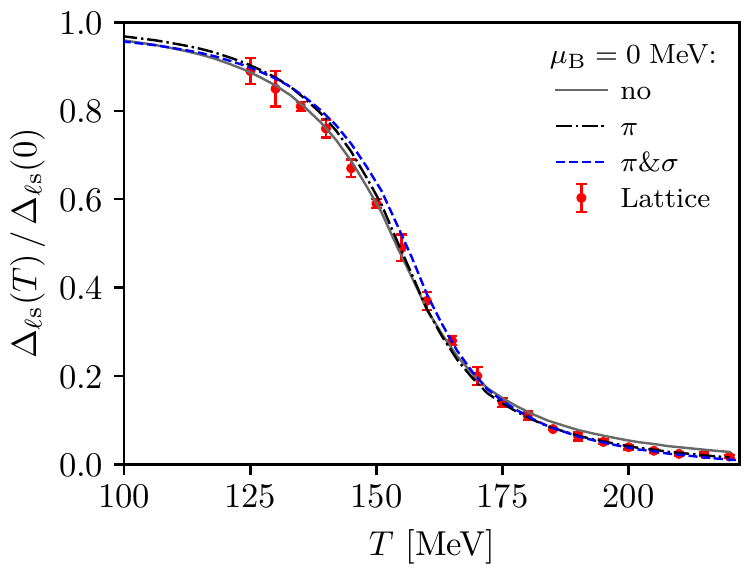}
\caption{\label{fig:norm_reg_condensat_comparison_meson_bc}Vacuum-normalized regularized quark condensate of the truncated DSE calculation at $\muB=0$ without meson backcoupling (solid, gray line; taken from Ref.~\cite{Gunkel:2019xnh}) and with pion (dashed-dotted, black line) as well as pion and sigma (dashed, blue line) backcoupling compared to continuum-extrapolated lattice results \cite{Borsanyi:2010bp} (solid, red circles). The parameters for the meson backcoupling data were rescaled as detailed in the text.}
\end{figure}

To study the impact of the backcoupled mesons individually, we consider the dynamical quark mass first in the vacuum and display 
corresponding results in Fig.~\ref{fig:bc_dependence_quark_df} for different sets of backcoupled mesons. 
Compared to the calculation without backcoupling, the inclusion of pionic backcoupling reduces the 
dynamical quark mass by around \SI{7}{\percent} at vanishing momenta. The sigma meson leads to a comparable reduction of chiral symmetry breaking of 
about $\SI{5}{\percent}$. Together, the effects add up to about $\SI{12}{\percent}$ reduction. The corresponding wave function renormalization function $Z_\textup{F}^f(p^2) = 1 / A_f(p^2)$ reacts 
much stronger to the pion backcoupling as to the sigma-meson backcoupling as shown in Fig.~\ref{fig:bc_dependence_quark_df}.

Next we switch on temperature and study the effect on the shape of the condensate as a function of $T$. We extract the pseudocritical 
temperature of the chiral crossover at zero chemical potential from the inflection point of the regularized quark condensate, i.e.,
\begin{align}
	\label{eq:pseudo_critical_temperature}
	\Tc(\muB)=\argmax_T\left\lvert\frac{\partial\Delta_{\ell\ups}(T,\muB)}{\partial T}\right\rvert .
\end{align}
In order to gauge the strength of our ansatz for the quark-gluon vertex, Eq.~\eqref{vertexa}, we adapt the parameter $d_1^{\+\upq}$ in the
quark DSE such that in all
considered cases we obtain the same pseudocritical transition temperature matched to the results from lattice QCD
\cite{Borsanyi:2010bp,Bazavov:2018mes}. To be precise, purely for reasons of better visibility in the plot we matched the setup 
without mesons and the setup with only pion backcoupling to the pseudocritical temperature of Ref.~\cite{Borsanyi:2010bp} and the 
full result with $\pi\&\sigma$ backcoupling to the one of Ref.~\cite{Bazavov:2018mes}, which is about 2 MeV larger. Since the numerical 
error of our matching procedure is anyhow about the same size as this difference there is no significance attached. The resulting values 
are given in Tab.~\ref{tab:truncation_parameter} together with the resulting pseudocritical temperature at vanishing chemical potential.

\begin{table}[t!]
\begin{ruledtabular}
\caption{\label{tab:truncation_parameter}Parameters of the truncation and parameters of QCD together with the pseudocritical temperature at vanishing chemical potential. The first two entries are the 
	vertex-strength parameters $d_1^{\+\textup{q}}$ and $d_1$ used in the quark and the gluon DSE, respectively. $m_\upq^\ell$ and 
	$m_\upq^\ups$ are the light and strange quark masses at an renormalization point of $\SI{80}{\GeV}$. The last entry is the resulting pseudocritical temperature at vanishing chemical potential.}
\begin{tabular}{cccccc}
&&&&&\\[-0.85em]
& $d_{1}^{\+\upq}$ [\si{\GeV\squared}] & $d_1$ [\si{\GeV\squared}] & $m_\upq^\ell$ [\si{\MeV}] & $m_\upq^\ups$ [\si{\MeV}] & $\Tc(0)$ [\si{\MeV}] \\[0.25em]
\hline
&&&&&\\[-0.75em]
no         & 12.85 & 8.49 & 1.47 & 37.8 & 155\,$\pm$\,1 \\[0.25em]
$\pi$         & 13.54 & 8.49 & 1.47 & 37.8 & 156\,$\pm$\,1 \\[0.25em]
$\pi\&\sigma$ & 14.18 & 8.49 & 1.47 & 37.8 & 157\,$\pm$\,1 \\[0.15em]
\end{tabular}
\end{ruledtabular}
\end{table}

In Fig.~\ref{fig:norm_reg_condensat_comparison_meson_bc}, the vacuum-normalized regularized quark condensate is plotted against the 
temperature at vanishing chemical potential for the two rescaled parameter sets with meson backcoupling. We additionally compare 
with corresponding lattice data from Ref.~\cite{Borsanyi:2010bp} and previous data without meson backcoupling from Ref.~\cite{Gunkel:2019xnh}.
Within error bars, all setups agree well with the lattice data. Note that the systematic shift of the full setup ($\pi\&\sigma$) of about 
\SI{2}{\MeV} to the right is a trivial result of the slightly larger transition temperature this set is matched to. For large temperatures,
the sets with meson backcoupling agree with each other and with the lattice data, whereas the set without backcoupling reveals slightly too large quark masses, which we did not fine-tune for simplicity.

% ==============================================================================
\section{\label{sec:effect_QCD_chiral_phase_diagram}Effect on the QCD chiral phase diagram}
% ==============================================================================

\begin{figure}[t!]
\centering
\includegraphics[width=\linewidth]{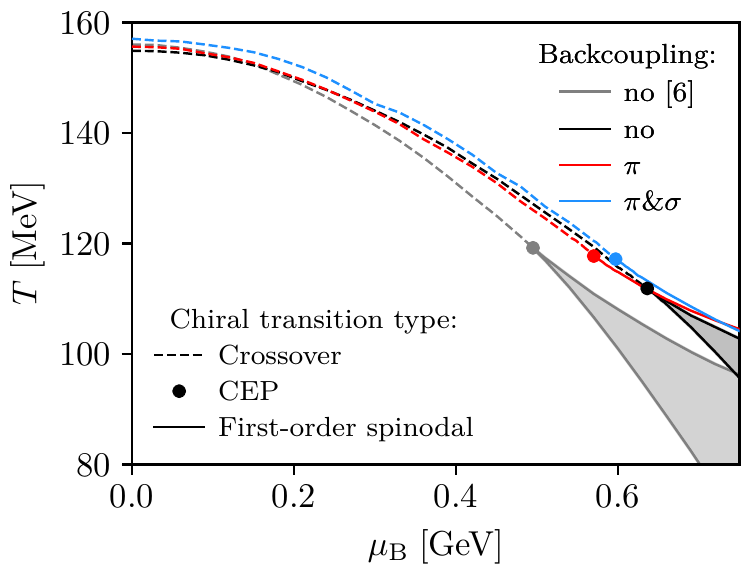}
\caption{\label{fig:phasediagram} QCD phase diagram for the setups explained in the text: no meson contributions to the quark-gluon interaction
	(black), taking into account only the pion backcoupling (red) and both, pion and sigma (blue). We further compare with results from Ref.~\cite{Isserstedt:2019pgx} (gray). Dashed lines correspond to crossover 
	transitions while solid lines represent a first-order spinodal. The big dots show the location of the second-order CEP of the corresponding truncation. The shaded area represents the coexistence region within the physical first-order phase transition takes place.}
\end{figure}

In Fig.~\ref{fig:phasediagram}, we show the chiral symmetry QCD phase diagram for the two rescaled meson backcoupling parameter sets 
introduced above and $\Nf=2+1$ quark flavors. We compare these two sets with the corresponding phase 
diagram without meson backcoupling (first results for this case have been reported in a contribution to conference proceedings \cite{Gunkel:2019enr}) and previous results from Ref.~\cite{Isserstedt:2019pgx}.
In each case, we find a crossover at low chemical potential that becomes steeper for increasing chemical potential and terminates in a second-order CEP followed by the coexistence region of a first-order transition. This coexistence region 
is bound by the spinodals of the chirally-broken Nambu--Goldstone and the chirally-symmetric Wigner--Weyl solution of the quark DSE. For the 
truncations with meson backcoupling, we only show the upper boundary.\footnote{The reason is technical: a determination of the lower boundary 
would require the calculation of the bound-state amplitudes using the Wigner--Weyl solutions. This is technically demanding and has, so far,
only been accomplished in the vacuum \cite{Hilger:2015zva}.}

\begin{table}[t!]
	\begin{ruledtabular}
		\caption{\label{tab:CEP_curvature}CEP and curvature $\kappa$ of the parameter sets introduced above with meson backcoupling and corresponding values for the truncations without meson backcoupling.}
		\begin{tabular}{ccccc}
			&& \\[-0.75em]
			Backcoupling  & $(\muB, T)_\textup{CEP}$ [\si{\MeV}] & $\kappa$ \\[0.25em]
			\hline
			&& \\[-0.75em]
			no \cite{Isserstedt:2019pgx}  & $(495, 119)$ & $0.0246$ \\[0.25em]
			no  & $(636, 112)$ & $0.0173$ \\[0.25em]
			$\pi$  & $(570, 118)$ & $0.0210$ \\[0.25em]
			$\pi\&\sigma$  & $(600, 117)$ & $0.0167$ \\[0.25em]
		\end{tabular}
	\end{ruledtabular}
\end{table}

First, we need to discuss the shift in the CEP not associated with the meson backcoupling but with the additional approximation
Eq.~\eqref{vertexa} as compared to the more advanced truncation using Eq.~(\ref{vertex1}) for both, the quark-gluon
interaction in the gluon and the quark DSE (gray lines in Fig.~\ref{fig:phasediagram} from Ref.~\cite{Fischer:2012vc,Fischer:2014ata,Isserstedt:2019pgx}). 
Comparing the two CEPs for the truncations without backcoupling in Tab.~\ref{tab:CEP_curvature}, we 
observe that the main effect of the additional approximation is a (considerable) shift of the CEP to larger chemical potential 
by almost \SI{30}{\percent}. This large shift emphasizes the need to carefully taking into account temperature and chemical potential effects 
in the leading structure of the quark-gluon vertex, as it has been done in previous work. 
Thus, the absolute values for the location of the CEP presented in this work should not be regarded as best results available
but only serve to highlight the relative difference of calculations suppressing and including the explicit influence of 
meson effects.

As a result, we find that the meson backcoupling effects on the quark (and the associated additional terms generated in the 
quark-gluon interaction) have only a small effect on the location of the CEP. The most prominent effect 
of the introduction of the mesonic backcoupling is the shift of the CEP towards (slightly) lower chemical potential and 
(slightly) higher temperatures. We find a chemical-potential shift of $\SI{10}{\percent}$ for pion and a reduced shift of only $\SI{6}{\percent}$ 
for pion and sigma meson backcoupling as compared to the result with no backcoupling. The location of the three CEPs are detailed 
in Tab.~\ref{tab:CEP_curvature}.

Overall, this is similar to the results found in Ref.~\cite{Eichmann:2015kfa} for effects due to baryon backcoupling.\footnote{Note, 
	however, that the study here is technically more advanced: whereas in Ref.~\cite{Eichmann:2015kfa} 
	only vacuum BSAs for the baryons have been taken into account, here we work with the full chemical-potential dependence of 
	their BSAs as determined in \cite{Gunkel:2019xnh,Gunkel:2020wcl}.} We find strong evidence that the location of the CEP is 
mainly driven by the nonresonant part of the quark-gluon vertex and the gluon, i.e., by the microscopic degrees 
of freedom of QCD. Of course, this is not true for other properties of the CEP like its critical exponents. These are expected 
to be driven by the long range degrees of freedom, in our case the sigma meson, in accordance with the expected $\ZZ(2)$ universality 
class of the Ising model in three dimensions \cite{Schaefer:2006ds}. A corresponding analytic scaling analysis 
in our framework is straightforward along the lines of Ref.~\cite{Fischer:2011pk}---the numerical confirmation, however, would require 
a tremendous additional effort that is outside the scope of this work.

Finally, we observe changes in the curvature of the crossover line at small chemical potentials. The curvature $\kappa$ 
is defined by
\beq
\label{eq:curvature}
\frac{\Tc(\muB)}{\Tc(0)}=1-\kappa\left(\frac{\muB}{\Tc(0)}\right)^2+\ldots \, .
\eeq
We find an increase of the curvature with the introduction of pion backcoupling effects and a slight decrease when taking into
account both, pion and sigma backcoupling effects. The corresponding values are shown in Tab.~\ref{tab:CEP_curvature}. The biggest influence on the curvature, however, has the additional approximation of Eq.~\eqref{vertexa} as can clearly be seen in the QCD phase diagram.

% ==============================================================================
\section{\label{sec:summary} Conclusions and outlook}
% ==============================================================================

In this work, we studied the effect of (off-shell) meson contributions to the quark-gluon vertex onto the location of the 
CEP of QCD as determined by functional methods from a coupled set of DSEs. Our 
study suggests that these effects are qualitatively irrelevant and quantitatively small. 
%\cPI{Is this in disagreement with FRG results? Heidelberg always emphasizes the importance of 
%their dynamical hadronization and so on...} 
The location of the CEP is driven
to a large extent by the microscopic degrees of freedom of QCD, the quarks and gluons. This is the main results of the present work
and agrees with previous findings for effects due to the backcoupling of baryons onto the quarks \cite{Eichmann:2015kfa}.

It should be kept in mind, however, that this is only true with regard to the location of the CEP. With regard to its properties, 
in particular with regard to the critical behavior very close to the CEP, it is expected that macroscopic degrees of freedom 
(in particular the sigma meson) take over as expected from a system in the $\ZZ(2)$ universality class. This has been explored in detail 
in effective models of QCD (see, e.g., Refs.~\cite{Schaefer:2006ds,Schaefer:2011ex,Chen:2021iuo}). A corresponding analysis in our framework requires 
additional efforts and is postponed to future work.

% ==============================================================================

\begin{acknowledgments}
	We thank Philipp Isserstedt, Jan Pawlowski and Bernd-Jochen Schaefer for 
	valuable discussions and a careful reading of the manuscript. 
	We are furthermore grateful to Richard Williams for valuable discussions.
	This work has been supported by the Helmholtz Graduate School for Hadron and Ion Research
	(HGS-HIRe) for FAIR, the GSI Helmholtzzentrum f\"{u}r Schwerionenforschung,
	and the BMBF under contract no.~05P18RGFCA.
\end{acknowledgments}

% ==============================================================================
\appendix*
% ==============================================================================

% ==============================================================================

\bibliography{meson_bc}
	
\end{document}